\documentclass[preprint]{ptephy_v1}

\preprintnumber{arXiv:2407.21784, YITP-24-95} 
\usepackage{hyperref}

\usepackage[all]{xy}
\usepackage{tcolorbox}
\usepackage{tcolorbox}
\tcbuselibrary{breakable, skins, theorems}

\newtheorem{theorem}{Theorem}

\newtheorem{definition}{Definition}
\newtheorem{remark}{Remark}
\newtheorem*{theorem*}{Theorem}

\numberwithin{equation}{subsection}
\numberwithin{definition}{subsection}
\numberwithin{theorem}{subsection}
\numberwithin{remark}{subsection}

\begin{document}

\title{A concrete construction of a topological operator in factorization algebras}


\author{Masashi Kawahira}
\affil{Yukawa Institute for Theoretical Physics, Kyoto University, Kitashirakawa Oiwakecho, Sakyo-ku, Kyoto 606-8502 Japan \email{masashi.kawahira@yukawa.kyoto-u.ac.jp}}





\begin{abstract}%
Factorization algebras play a central role in the formulation of quantum field theories given by Kevin Costello and Owen Gwilliam. In this paper, we propose a concrete construction of a topological operator in their formulation. 
We focus on a shift symmetry of a one-dimensional massless scalar theory. And we give some discussions of $\mathbb{Z}$-gauging of it, i.e., compact scalar theory and its $\theta$-vacuum.
\end{abstract}

\subjectindex{A12, A13, B05, B34, B39}

\maketitle

\tableofcontents

\newpage
\section{Introduction}
Quantum field theory is a central topic in theoretical physics. This is valid to describe various kinds of phenomena and also useful to generate some mathematical conjectures.
In spite of such effectiveness, we have not understood the complete mathematical formulation of quantum field theories.

In the last ten years, Kevin Costello and Owen Gwilliam have developed a new formulation which is based on factorization algebras\cite{Costello:2016vjw}\cite{Costello:2021jvx}.
This formulation works well in perturbative field theories, conformal field theories and topological quantum filed theories. These days, some people are working on formulations of non-perturbative aspects of quantum field theories \cite{Alfonsi:2023qpv}\cite{Elliott:2014haa}. 

In such a context, non-perturbative Noether theorem is also proposed by \cite{Costello:2023knl} which is motivated by recent developments of generalized symmetries \cite{Kapustin:2014gua}\cite{Gaiotto:2014kfa}\cite{Sharpe:2015mja}.
The dogma of generalized symmetries is 
\begin{align}
    {\rm Symmetry}={\rm Topological\ operator}.
\end{align}
A topological operator with $\mathcal{U}(U)$ some support $U$ is \textit{invariant} under homotopic transformations of $U$.

A question arises: what does this \textit{invariance} mean ? 
We will show that $\mathcal{U}(U_1)$ and $\mathcal{U}(U_2)$ are in the same equivalence class of Batalin-Vilkovisky cohomology when $U_1$ and $U_2$ are homotopic.
We will propose that this is the definition of the topological operators.

In this paper, we focus only on the case of one-dimensional free massless scalar theory (i.e., quantum mechanics without potential terms) .
This theory has a shift symmetry $\Phi\mapsto\Phi+\alpha$ where $\Phi$ is a field and $\alpha$ is a real number.
We give a concrete construction of the topological operator representing this shift $\mathbb{R}$-symmetry.
A virtue of the topological operator is that we can perform gauging of the subsymmetry $\mathbb{Z}\subset \mathbb{R}$. 
By the $\mathbb{Z}$-gauging, we can obtain a compact scalar theory. It is an interesting theory because this theory has instanton effects. 
There is a work about compact $U(1)$ $p$-form gauge theory\cite{Elliott:2014haa} motivated by S-duality. 
The case of $p=0$ is essentially the same as this paper in terms of using $\mathbb{Z}$-gauging. However, our construction is more concrete, and we will also discuss the relation to generalized symmetry notions and $\theta$ vacuum.

This paper is organized as follows. 
In section 2, we will review the formulation by Costello and Gwilliam.
In section 3, we will propose a concrete construction of the topological operator.
In section 4, we will see the $\theta$-degree and the formulation.
Section 5 is devoted for conclusion and discussion.

\section{Review of factorization algebra of free real scalar theory}
\subsection{Observable space}
In this section, we will define an observable $\mathcal{O}$. Let $M$ be a $d$-dimensional manifold and a real scalar field $\Phi$ are in $C^\infty (M)$. 
$\mathcal{O}$ must have two properties:
\begin{itemize}
    \item $\mathcal{O}$ is a functional, i.e., $\mathcal{O}$ is a map from a field $\Phi$ to a number $\mathcal{O}(\Phi)$.
    \item $\mathcal{O}$ has a concept of \textit{locality}. In other words, $\mathcal{O}$ has support $U\subset M$.
\end{itemize}
Here $U$ is an open subset of $M$.
If the shape of $U$ is like a $d$-dimensional open ball $B^d$, $\mathcal{O}$ is a point operator with a UV cut-off. In the case of $U$ being $B^{d-1}\times S^1$, it is a loop operator.

Motivated by the above two properties, we will find an example of observables: \textit{linear observable}.
\begin{align}
    \mathcal{O}_{\rm linear}:\Phi\mapsto \int_M f\Phi,\ \ 
    f\in C_{\rm c}^\infty(U,\mathbb{C})
\end{align}
where $f$ is a $\mathbb{C}$-function with compact support in $U$. The integral $\int_M f\Phi$ refer to only the $\Phi$ on $U$, hence in this sense $\mathcal{O}_{\rm linear}$ has locality.
We identify $\mathcal{O}_{\rm linear}$ with $f$, then let define \textit{linear observable space on }$U$ as
\begin{align}
    {\rm Obs}_{\rm linear}(U):=C_{\rm c}^\infty(U,\mathbb{C}).
\end{align}

A more general observable can be regarded as a polynomial of functions with compact supports:
$\mathcal{O}=c+f+f_1*f_2+\cdots$,
where $*$ is a formal symmetric product, $c\in\mathbb{C}$ and $f,f_1,f_2,\cdots\in C_{\rm c}^\infty(U,\mathbb{C})$. And $\mathcal{O}$ acts $\Phi$ as
\begin{align}
    \mathcal{O}:\Phi\mapsto
    c
    +\int_M f\Phi
    +\int_M f_1\Phi \int_M f_2\Phi
    +\cdots
\end{align}
\begin{tcolorbox}[colframe=red,colback=red!3!]
\begin{definition}
Observable space ${\rm Obs}(U)$ is defined as 
\begin{align}
    {\rm Obs}(U):={\rm Sym}(C_{\rm c}^\infty(U,\mathbb{C}))
\end{align} and the elements are called observables on $U$.
\end{definition}
 \end{tcolorbox}
 
\subsection{Classical derived observable space}
Roughly speaking, \textit{derived} means that we will add a concept of \textit{degree} to the observable space ${\rm Obs}(U)$.
This formulation is originally given by Batalin and Vilkovisky\cite{Batalin:1981jr}\cite{Batalin:1983ggl}.
\begin{align}
    &C_{\rm c}^\infty(U)^{0}:=(C_{\rm c}^\infty(U,\mathbb{C}),0),\\
    &C_{\rm c}^\infty(U)^{-1}:=(C_{\rm c}^\infty(U,\mathbb{C}),-1).
\end{align}
The first one is a \textit{degree-0 linear observable}, and the second one is a \textit{degree-$(-1)$ linear observable} which corresponds to \textit{anti-field} in physics literature.
For simplicity, we denote $(f,0)\in C_{\rm c}^\infty(U)^{0}$ and $(f,-1)\in C_{\rm c}^\infty(U)^{-1}$ as $f$ and $f^\star$.
The symmetric product $*$ is defined for them as
\begin{align}
    a*b=(-1)^{|a||b|}b*a\label{eq:*}.
\end{align}
\begin{tcolorbox}[colframe=red,colback=red!3!]
\begin{definition}
    Classical derived observable space ${\rm Obs}^{\rm cl}(U)$ is defined as
    \begin{align}
        {\rm Obs}^{\rm cl}(U)
        :=
        \left({\rm Sym}(C_{\rm c}^\infty(U)^{-1}\oplus C_{\rm c}^\infty(U)^{0}),\Delta_{\rm BV}^{\rm cl}\right)
    \end{align}
    where $\Delta_{\rm BV}^{\rm cl}$ is a classical Batalin-Vilkovisky operator which is defined in Definition \ref{def:cl_BV_op}.
\end{definition}    
\end{tcolorbox}
By (\ref{eq:*}), we can rewrite classical derived observable space ${\rm Obs}^{\rm cl}(U)$.\footnote{Precisely, we need to take a completion in order to make ${\rm Obs}^{\rm cl}(U)$ a topological vector space. However we omit the discussion. If you are interested in it, check \cite{Costello:2016vjw}.\label{ref:completion}}
\begin{align}
  {\rm Obs}^{\rm cl}(U)  
  =\Bigg(\cdots 
  &\oplus
  \bigg(\bigwedge^2 C_{\rm c}^\infty(U)^{-1}* {\rm Sym}\left(C_{\rm c}^\infty(U)^{0}\right)\bigg)\notag\\
  &\oplus 
  \bigg(C_{\rm c}^\infty(U)^{-1}* {\rm Sym}\left(C_{\rm c}^\infty(U)^{0}\right)\bigg)\notag\\
  &\oplus
  {\rm Sym}\left(C_{\rm c}^\infty(U)^{0}\right)
  ,\Delta_{\rm BV}^{\rm cl}\Bigg).
\end{align}
This is a decomposition about the degree.
The first line is degree $(-2)$, second one is degree $(-1)$, and third one is degree $0$. Actually $\Delta_{\rm BV}^{\rm cl}$ is defined as a map from degree $(-n)$ to degree $(-n+1)$:
\begin{align}
    \Delta_{\rm BV}^{\rm cl}:
    \bigwedge^n C_{\rm c}^\infty(U)^{-1}* {\rm Sym}\left(C_{\rm c}^\infty(U)^{0}\right)
    \to
    \bigwedge^{n-1} C_{\rm c}^\infty(U)^{-1}* {\rm Sym}\left(C_{\rm c}^\infty(U)^{0}\right).
\end{align}
The concrete form of $\Delta_{\rm BV}^{\rm cl}$ depends on what theory we want. In this paper, we are interested in free theory, thus we will defined as follows.
\begin{tcolorbox}[colframe=red,colback=red!3!]
\begin{definition}\label{def:cl_BV_op}
     Classical Batalin-Vilkovisky operator $\Delta_{\rm BV}^{\rm cl}$ is a map $C_{\rm c}^\infty(U)^{-1}\ni f^\star\mapsto -(-\Delta+m^2)f\in C_{\rm c}^\infty(U)^{0}$. Here $\Delta$ is a Laplacian of $M$. And we define it with Leibniz rule, then we have
     \begin{align}
    \Delta_{\rm BV}^{\rm cl}(f_1^{\star}*\cdots *f_n^{\star}* P)
    &=\sum_{i=1}^n f_1^{\star}*\cdots*f_{i-1}^{\star}* (-1)^{i-1}(\Delta_{\rm BV}^{\rm cl} f_i^{\star}) *f_{i+1}^{\star}*\cdots*f_n^{\star}*P\notag\\
    &=\sum_{i=1}^n f_1^{\star}*\cdots* \widehat{f_i^{\star}} *\cdots*f_n^{\star}*(-1)^{i-1}(\Delta_{\rm BV}^{\rm cl} f_i^{\star})*P
    \end{align}
    where $f_i^{\star}\in C_{\rm c}^\infty(U)^{-1},\ P\in{\rm Sym}\left(C_{\rm c}^\infty(U)^{0}\right)$.
    And the $(-1)^{i-1}$ comes from the rule that $\Delta_{\rm BV}^{\rm cl}$ and $f_i^\star$ are anti-commutative, since $\Delta_{\rm BV}^{\rm cl}$ has a degree $+1$.
\end{definition}    
\end{tcolorbox}

By some calculations, we have 
\begin{itemize}
    \item \textit{Leibniz rule} : $\Delta_{\rm BV}^{\rm cl}(A*B)
    =
    \Delta_{\rm BV}^{\rm cl}(A)*B
    +
    (-1)^{|A|}A*\Delta_{\rm BV}^{\rm cl}(B)$,
    \item \textit{Nilpotency} : $\left(\Delta_{\rm BV}^{\rm cl}\right)^2=0$.
\end{itemize}
\textit{Nilpotency} means that
\begin{align}
   &{\rm Obs}^{\rm cl}(U)= 
   \notag\\
   &\bigg(\cdots 
   \xrightarrow{\Delta_{\rm BV}^{\rm cl}}\bigwedge^2 C_{\rm c}^\infty(U)^{-1}* {\rm Sym}\left(C_{\rm c}^\infty(U)^{0}\right)
   \xrightarrow{\Delta_{\rm BV}^{\rm cl}}C_{\rm c}^\infty(U)^{-1}* {\rm Sym}\left(C_{\rm c}^\infty(U)^{0}\right)
   \xrightarrow{\Delta_{\rm BV}^{\rm cl}}{\rm Sym}\left(C_{\rm c}^\infty(U)^{0}\right)\bigg)
\end{align}
is a chain complex. This is called \textit{classical Batalin–Vilkovisky complex}.
The cohomology of classical Batalin–Vilkovisky complex
 is called \textit{classical Batalin–Vilkovisky cohomology}:
 \begin{align}
   H^*\left({\rm Obs}^{\rm cl}(U)\right).
 \end{align}
\textit{Leibniz rule} means that $H^*\left({\rm Obs}^{\rm cl}(U)\right)$ has a product $*$, because the product of $A+\Delta_{\rm BV}^{\rm cl}(\cdots)$ and $B+\Delta_{\rm BV}^{\rm cl}(\cdots)$ can be rewritten as $A*B+\Delta_{\rm BV}^{\rm cl}(\cdots)$.
 \footnote{To check this, let us think about $ (A+\Delta_{\rm BV}^{\rm cl} \tilde{A})*(B+\Delta_{\rm BV}^{\rm cl} \tilde{B})$ where $A,B$ satisfy $\Delta_{\rm BV}^{\rm cl}A=0,\ \Delta_{\rm BV}^{\rm cl}B=0$. 
   \begin{align}
     (A+\Delta_{\rm BV}^{\rm cl} \tilde{A})*(B+\Delta_{\rm BV}^{\rm cl} \tilde{B}) 
     &=
     A*B+(\Delta_{\rm BV}^{\rm cl}\tilde A)*B+A*(\Delta_{\rm BV}^{\rm cl}\tilde B)+(\Delta_{\rm BV}^{\rm cl}\tilde A)*(\Delta_{\rm BV}^{\rm cl}\tilde B)\notag\\
     &=
     A*B  
     +(\Delta_{\rm BV}^{\rm cl}\tilde A)*B+(-1)^{|\tilde{A}|}\tilde A*(\Delta_{\rm BV}^{\rm cl}B)  
     +A*(\Delta_{\rm BV}^{\rm cl}\tilde B)\notag\\
     &+(-1)^{|A|}(\Delta_{\rm BV}^{\rm cl}A)*\tilde{B}  
     +(\Delta_{\rm BV}^{\rm cl}\tilde A)*(\Delta_{\rm BV}^{\rm cl}\tilde B)
     +(-1)^{|\tilde{A}|}\tilde{A}*(\Delta_{\rm BV}^{\rm cl})^2\tilde{B}\notag\\
     &=
     A*B  +\Delta_{\rm BV}^{\rm cl}(\tilde A*B)  +(-1)^{|A|}\Delta_{\rm BV}^{\rm cl}(A*\tilde B) +\Delta_{\rm BV}^{\rm cl}(\tilde A*(\Delta_{\rm BV}^{\rm cl}\tilde B)) \notag\\
     &=A*B+\Delta_{\rm BV}^{\rm cl}(\cdots)
   \end{align}}
\begin{remark}
   On the other hand, quantum Batalin-Vilkovisky operator $\Delta_{\rm BV}^{\rm q}$ does NOT have Leibniz rule, as we discuss later, then quantum cohomology $H^*\left({\rm Obs}^{\rm q}(U)\right)$ does NOT have product $*$.
 \end{remark}

Let us consider the physical meanings of the cohomology $H^0\left({\rm Obs}^{\rm cl}(U)\right)$.
We take a 0-degree observables $\mathcal{O}_1,\mathcal{O}_2$ and assume that these are same in the cohomology, i.e., $\exists X\ {\rm s.t.}$
\begin{align}
    \mathcal{O}_2-\mathcal{O}_1
    =
    \Delta_{\rm BV}^{\rm cl}X.
\end{align}
Let $\Phi_{\rm cl}$ be a solution of the equation of motion $(-\Delta+m^2)\Phi=0$\footnote{If you want a theory with some interactions, we need to add some terms to $\Delta_{\rm BV}^{\rm cl}$.}. We have
\begin{align}
    \mathcal{O}_2(\Phi_{\rm cl})-\mathcal{O}_1(\Phi_{\rm cl})
    &=
    \Delta_{\rm BV}^{\rm cl}X(\Phi_{\rm cl})\notag\\
    &=
    0.
\end{align}
Hence we can see $H^0\left({\rm Obs}^{\rm cl}(U)\right)$ as the on-shell evaluation of observables. However $[\mathcal{O}_1](=[\mathcal{O}_2])$ is not a number. Then we define a map called \textit{state} $\langle-\rangle$.
\begin{tcolorbox}[colframe=red,colback=red!3!]
\begin{definition}
    A state $\langle-\rangle$ is a smooth map:
    \begin{align}
        \langle-\rangle:
        H^0\left({\rm Obs}^{\rm cl}(M)\right)
        \to 
        \mathbb{C}.
    \end{align}
\end{definition}
\end{tcolorbox}
\subsection{Concrete calculations of classical Batalin-Vilkovisly cohomology}
In some situation, we can calculate $H^*\left({\rm Obs}^{\rm cl}(M)\right)$ explicitly.
\begin{tcolorbox}[colframe=blue,colback=blue!3!]
\begin{theorem}\label{thm:M_compact_massive_cl}
    If $M$ is compact and $m^2>0$, then
    \begin{align}
        H^n\left({\rm Obs}^{\rm cl}(M)\right)
        =
        \left\{
        \begin{array}{ll}
        \mathbb{C} & (n=0) \\
        0 & ({\rm otherwise})
        \end{array}
        \right.
    \end{align}
\end{theorem}
\end{tcolorbox}
\begin{proof}
    \begin{align}
        A:=
        \left(C_{\rm c}^\infty(M)^{-1}\xrightarrow{\Delta_{\rm BV}^{\rm cl}}C_{\rm c}^\infty(M)^0\right)
    \end{align}
    This is an isomorphism. Thus $H^*(A)=0$, then $H^0({\rm Sym}(A))=\mathbb{C}$ and $H^{n\le -1}({\rm Sym}(A))=0$
\end{proof}

\begin{tcolorbox}[colframe=blue,colback=blue!3!]
\begin{theorem}\label{thm:M=I_cl}
    If $M=\mathbb{R}$ and $I\subset M$ is an interval, 
    \begin{align}
        H^n\left({\rm Obs}^{\rm cl}(I)\right)
        =
        \left\{
        \begin{array}{ll}
        \mathbb{C}[q,p] & (n=0) \\
        0 & ({\rm otherwise})
        \end{array}
        \right.
    \end{align}
    Here $q,p$ has degree 0.
\end{theorem}
\end{tcolorbox}
\begin{proof}
    We show a quasi-isomorphism:
    \begin{align}
        A\sim B
    \end{align}
    where
    \begin{align}
        A=\left(C_{\rm c}^\infty(I)^{-1}\xrightarrow{\Delta_{\rm BV}^{\rm cl}}C_{\rm c}^\infty(I)^0\right),\ 
        B=(0\to\mathbb{C}^2).
    \end{align}
    $\mathbb{C}^2$ sits in degree 0. And we denote the basis of $\mathbb{C}^2$ as $q,p$.
    The cohomology of ${\rm Sym}(A)$ and ${\rm Sym}(B)$ are $H^*({\rm Obs}^{\rm cl}(I))$ and $\mathbb{C}[q,p]$ respectively.
    
    First of all, we will see the following commutative diagram:
    \begin{align}
        \xymatrix{
    C_{\rm c}^\infty(I)^{-1} \ar[r] \ar[d]_{\pi_1} & C_{\rm c}^\infty(I)^0 \ar[d]_{\pi_2} \\
    0 \ar[r] & \mathbb{C}^2
  }
    \end{align}
    where $\pi_2$ is defined as  
   \begin{align}
       \pi_2(g)
       :=q\int_I {\rm d}x\ g(x)\phi_q(x)+p\int_I {\rm d}x\ g(x)\phi_p(x)
   \end{align}
   for $g\in C_{\rm c}^\infty(I)^0$.
   The definition of $\phi_q,\phi_p\in C^\infty(\mathbb{R})$ is as follows.
    In case that $m>0$ we define $\phi_q,\phi_p\in C^\infty(\mathbb{R})$ as
\begin{align}
    \phi_q(x)=\frac{1}{2}(e^{mx}+e^{-mx}),\ 
    \phi_p(x)=\frac{1}{2m}(e^{mx}-e^{-mx}).
\end{align}
They form the kernel of $-\Delta+m^2$.
If $m=0$, we define
\begin{align}
    \phi_q(x)=1,\ 
    \phi_p(x)=x.
\end{align}
We note that $\phi_p'(x)=\phi_q(x)$.
We can easily check that
   $\pi_2(g)=0$ holds if $g=\Delta_{\rm BV}^{\rm cl}f^{\star}.$

Next, we will see $H^0(A)=H^0(B)$. By definition,
\begin{align}
    H^0(A)
    =
    \frac{C_{\rm c}^\infty(I)^0}
    {{\rm im}(\Delta_{\rm BV}^{\rm cl})}.
\end{align}
We will show that ${\rm im}(\Delta_{\rm BV}^{\rm cl})={\rm ker}(\pi_2)$. If it holds, $H^0(A)=H^0(B)$. Obviously,
\begin{align}
    {\rm im}(\Delta_{\rm BV}^{\rm cl})\subset {\rm ker}(\pi_2).
\end{align}
Then we will check ${\rm im}(\Delta_{\rm BV}^{\rm cl})\supset {\rm ker}(\pi_2)$.
Take $f\in {\rm ker}(\pi_2)$. $f$ satisfies
\begin{align}
    &\int_I {\rm d}x\ f(x)e^{mx}=0\ {\rm and}
    \int_I {\rm d}x\ f(x)e^{-mx}=0\ \ 
    ({\rm for\ massive\ case}),\notag\\
    &\int_I {\rm d}x\ f(x)x=0\ {\rm and}
    \int_I {\rm d}x\ f(x)=0\ \ 
    ({\rm for\ massless\ case}) \label{eq:integral_of_f},
\end{align}
because $\int_I {\rm d}x f(x)\phi_p(x)=0$ and $\int_I {\rm d}x f(x)\phi_q(x)=0$.
Let $G\in C^0(\mathbb{R})$ be the Green function:
\begin{align}
    &G(x)=\frac{1}{2m}e^{-m|x|}\ \ 
    ({\rm for\ massive\ case}),\notag\\
    &G(x)=-\frac{1}{2}|x|\ \ 
    ({\rm for\ massless\ case}).
\end{align}
Then the convolution of $f$ and $G$ is
\begin{align}
    (G\cdot f)(x):=\int_I {\rm d}y\ G(x-y)f(y).
\end{align}
This is in $C_{\rm c}^\infty(I)^0$ because of  (\ref{eq:integral_of_f}). For $(G\cdot f)^\star\in C_{\rm c}^\infty(I)^{-1}$, 
\begin{align}
    f=\Delta_{\rm BV}^{\rm cl}(G\cdot f)^\star.
\end{align}
Thus $f\in {\rm im}(\Delta_{\rm BV}^{\rm cl})$, and ${\rm im}(\Delta_{\rm BV}^{\rm cl})\supset {\rm ker}(\pi_2)$.

Finally, we will see $H^{-1}(A)=H^{-1}(B)$. Clearly $H^{-1}(B)=0$. Then let us think of
\begin{align}
    H^{-1}(A)={\rm ker}(\Delta_{\rm BV}^{\rm cl}).
\end{align}
This is trivial since there are not non-trivial solutions of 
\begin{align}
    (-\Delta+m^2)f=0
\end{align}
for $f\in C_{\rm c}^\infty(I)$. Then $H^{-1}(A)=H^{-1}(B)=0$.
\end{proof}

\begin{tcolorbox}[colframe=blue,colback=blue!3!]
\begin{theorem}\label{thm:M=R_cl}
    If $M=\mathbb{R}$, then for any value of $m$
    \begin{align}
        H^n\left({\rm Obs}^{\rm cl}(\mathbb{R})\right)
        =
        \left\{
        \begin{array}{ll}
        \mathbb{C}[q,p] & (n=0) \\
        0 & ({\rm otherwise})
        \end{array}
        \right.
    \end{align}
    Here $q,p$ has degree 0.
\end{theorem}
\end{tcolorbox}
\begin{proof}
    The proof of this is essentially same as Theorem \ref{thm:M=I_cl}.
\end{proof}

Theorem \ref{thm:M_compact_massive_cl} tells us that the state $\langle-\rangle$ can be given as an isomorphism, i.e., the state $\langle-\rangle$ is canonically determined. 
On the other hand, the case of Theorem \ref{thm:M=I_cl} and Theorem \ref{thm:M=R_cl}, there is room to choose the state $\langle-\rangle$. Why is there such a difference ? In massive case, we have a reasonable interpretation for this difference.

In the case of non-compact manifold $M_{\rm non}$, the boundary condition is necessary to compute the path integral.
If not, the scalar field $\Phi$ might diverge on $\partial \overline{M_{\rm non}}$.
Then the path integral calculation fails.
In order to compute the value of the path integral, we must choose a boundary condition.
Reflecting it, we must choose a map $\langle-\rangle:H^0\left({\rm Obs}^{\rm cl}(M_{\rm non})\right)\to\mathbb{C}$.

On the other hand, in the case of compact manifold $M_{\rm com}$, the scalar field $\Phi$ never diverge on $\partial M_{\rm com}$.
Thus we can decide the value of the path integral.
Similarly, we find the state $\langle-\rangle$ canonically.

The above discussion is valid for massive case.
However, in massless case, even for a compact manifold, $H^0\left({\rm Obs}^{\rm cl}(M_{\rm com})\right)$ is not isomorphic to $\mathbb{C}$. In other words, we can not decide the expectation value naturally, though there is no $\Phi$'s divergence on the boundary $\partial M_{\rm com}$.
In order to understand it, we have to see another factor to interrupt the path integral computation: \textit{IR divergence}. 
We will propose a treatment of IR divergence in \cite{Kawahira:2024}.

\subsection{Definition of position observable $Q$ and momentum observable $P$}\label{subsec:Q_P}
We define a position observable $Q$ as a function in $C_{\rm c}^\infty(I)$ satisying
\begin{align}
    \int_{I} Q(x)\phi_q(x)\ {\rm d}x=1
    \ {\rm and}\ 
    Q(-x)=Q(x).
\end{align}
By the second condition, we obtain $\int_{I} Q(x)\phi_p(x)\ {\rm d}x=0$.
And we define a momentum observable as
\begin{align}
    P:=-Q'\in C_{\rm c}^\infty(I)^{0}.
\end{align}

Now we see the reason why we call $Q$ and $P$ as position and momentum. If $Q,P$ act on field $\Phi\in C^\infty(\mathbb{R})$, we get
\begin{align}
    Q(\Phi)=\int_{I_t} Q(x)\Phi(x)\ {\rm d}x,\ 
    P(\Phi)=\int_{I_t} Q(x)\Phi'(x)\ {\rm d}x.
\end{align}
Especially in massless case, $Q(\cdot)$ is a ``smeared $\delta$-function'' 
because  we have
\begin{align}
    \int_{I} Q(x)\ {\rm d}x=1
    \ {\rm and}\ 
    Q(-x)=Q(x)
\end{align}
by using $\phi_q=1$.
Hence if we narrow the width of the interval, $Q$ get close to $\delta$-function, and $Q(\Phi)$ and $P(\Phi)$ approaches $\Phi(t)$ and $\Phi'(t)$.

In the massive case, $Q$ is also a kind of smeared $\delta$-functions. To see this, remind 
\begin{align}
    \phi_q(x)=\frac{1}{2}(e^{mx}+e^{-mx}).
\end{align}
Then we have
\begin{align}
    \int_I\delta(x)\phi_q(x)=1
    \ {\rm and}\ 
    \delta(-x)=\delta(x).
\end{align}
Therefore also in the massive case, we expect $Q$ get close to $\delta$-function.

And $Q$ and $P$ have the following important property.
\begin{tcolorbox}[colframe=blue,colback=blue!3!]
\begin{theorem}\label{thm:generator}
   $[Q]$ and $[P]$ are the generators of $H^0({\rm Obs}^{\rm cl}(\mathbb{R}))$
\end{theorem}
\end{tcolorbox}
\begin{proof}
    We have already seen $H^0({\rm Obs}^{\rm cl}(\mathbb{R}))=\mathbb{C}[q,p]$ in Theorem \ref{thm:M=R_cl}.
    Take
    \begin{align}
        Q+\Delta_{\rm BV}^{\rm cl}X, 
        P+\Delta_{\rm BV}^{\rm cl}Y,
    \end{align}
    where $X,Y\in {\rm Sym}(C_{\rm c}^\infty(\mathbb{R}))*C_{\rm c}^\infty(\mathbb{R}))^{-1}$.
    By $\pi_2$ action,
    \begin{align}
        \pi_2\left(Q+\Delta_{\rm BV}^{\rm cl}X\right)
        =
        \pi_2(Q)=q,\\
        \pi_2\left(P+\Delta_{\rm BV}^{\rm cl}Y\right)
        =
        \pi_2(P)=p.
    \end{align}
    We used $\phi'_p(x)=\phi_q(x)$.
\end{proof}

\subsection{Quantum derived observable space}
\begin{tcolorbox}[colframe=red,colback=red!3!]
\begin{definition}
    Quantum derived observable space ${\rm Obs}^{\rm q}(U)$ is defined as
    \begin{align}
        {\rm Obs}^{\rm q}(U)
        :=
        \left({\rm Sym}(C_{\rm c}^\infty(U)^{-1}\oplus C_{\rm c}^\infty(U)^{0}\oplus\mathbb{C}\hbar),\Delta_{\rm BV}^{\rm q}=\Delta_{\rm BV}^{\rm cl}+ \hbar\delta\right)
    \end{align}
    where $\Delta_{\rm BV}^{\rm q}=\Delta_{\rm BV}^{\rm cl}+ \hbar\delta$ is a quantum Batalin-Vilkovisky operator and $\delta$ is defined in Definition \ref{def:delta}.
\end{definition}    
\end{tcolorbox}
Note that $C_{\rm c}^\infty(U)*\mathbb{C}=C_{\rm c}^\infty(U)$, we can rewrite it as the formal series of $\hbar$:
\begin{align}
    {\rm Obs}^{\rm q}(U)= 
 \Bigg(\cdots 
 &\oplus\Bigg(\bigwedge^2 C_{\rm c}^\infty(U)^{-1}*{\rm Sym}\left(C_{\rm c}^\infty(U)^{0}\right)[\hbar]\Bigg)\notag\\
 &\oplus \Bigg(C_{\rm c}^\infty(U)^{-1}*{\rm Sym}\left(C_{\rm c}^\infty(U)^{0}\right)[\hbar]\Bigg)\notag\\
 &\oplus{\rm Sym}\left(C_{\rm c}^\infty(U)^{0}\right)[\hbar],\Delta_{\rm BV}^{\rm q}\Bigg). 
\end{align}
$\delta$ is a quantum correction and will be defined as a map from $-n$-degree to $-n+1$-degree:
\begin{align}
    \delta:
    \bigwedge^n C_{\rm c}^\infty(U)^{-1}* {\rm Sym}\left(C_{\rm c}^\infty(U)^{0}\right)
    \to
    \bigwedge^{n-1} C_{\rm c}^\infty(U)^{-1}* {\rm Sym}\left(C_{\rm c}^\infty(U)^{0}\right).
\end{align}
The definition of $\delta$ is a bit complicated. Then we take three steps.
\begin{tcolorbox}[colframe=red,colback=red!3!]
\begin{definition}\label{def:delta}
 
\mbox{ }

{\rm (1)} For $f^\star*g\in C_{\rm c}^\infty(U)^{-1}* C_{\rm c}^\infty(U)^{0}$, the quantum correction $\delta$ is defined as
\begin{align}
\begin{array}{ccc}
C_{\rm c}^\infty(U)^{-1}* C_{\rm c}^\infty(U)^{0} 
&\stackrel{\delta}{\longrightarrow} & 
\mathbb{R}\subset{\rm Sym}(C_{\rm c}^\infty(U)^0) \\
\rotatebox{90}{$\in$} & & \rotatebox{90}{$\in$} \\
f^\star*g & \longmapsto & \int_U f(x)g(x){\rm d}x
\end{array}
\end{align}
thus $ \delta$ has degree $+1$.

{\rm (2)} For $f^\star*P\in C_{\rm c}^\infty(U)^{-1}*{\rm Sym}\left(C_{\rm c}^\infty(U)^{0}\right)$, it is sufficient to think the $n$-th term $g_1*\cdots*g_n$ of $P$.
\begin{align}
\begin{array}{ccc}
C_{\rm c}^\infty(U)^{-1}*{\rm Sym}\left(C_{\rm c}^\infty(U)^{0}\right) 
&\stackrel{\delta}{\longrightarrow} & 
{\rm Sym}\left(C_{\rm c}^\infty(U)^{0}\right)\\
\rotatebox{90}{$\in$} & & \rotatebox{90}{$\in$} \\
f^\star*g_1*\cdots*g_n & \longmapsto & \sum_{i=1}^n g_1*\cdots*\widehat{g_i}*\cdots*g_n *\delta(f^\star*g_i)
\end{array}
\end{align}

{\rm (3)} For $f_1^\star*\cdots*f_m^\star*P\in
\bigwedge^m C_{\rm c}^\infty(U)^{-1}*{\rm Sym}\left(C_{\rm c}^\infty(U)^{0}\right)$
\begin{align}
\begin{array}{ccc}
\bigwedge^m C_{\rm c}^\infty(U)^{-1}*{\rm Sym}\left(C_{\rm c}^\infty(U)^{0}\right)
&\stackrel{\delta}{\longrightarrow} & 
\bigwedge^{m-1} C_{\rm c}^\infty(U)^{-1}*{\rm Sym}\left(C_{\rm c}^\infty(U)^{0}\right)
\\
\rotatebox{90}{$\in$} & & \rotatebox{90}{$\in$} \\
f_1^\star*\cdots*f_m^\star*P & \longmapsto & 
\sum_{i=1}^m f_1^\star*\cdots*\widehat{f_i^\star}*\cdots*f_m^\star*(-1)^{i-1}\delta(f_i^\star*P) .
\end{array}
\end{align}
\end{definition}
\end{tcolorbox}
$\delta$ is nilpotent\footnote{To check it, we use $\delta(f^\star*\delta(\tilde{f}^\star*P))=\delta(\tilde{f}^\star*\delta(f^\star*P))$.} and anti-commute with $\Delta_{\rm BV}^{\rm cl}$
\footnote{We can understand that it is because $\Delta_{\rm BV}^{\rm cl}$ and $\delta$ have degrees $+1$.}: $\delta^2=0,\ \Delta_{\rm BV}^{\rm cl}\delta=-\delta\Delta_{\rm BV}^{\rm cl}$.
Then we have
\begin{align}
   (\Delta_{\rm BV}^{\rm q})^2
   &=(\Delta_{\rm BV}^{\rm cl}+\hbar\delta)(\delta_{\rm BV}^{\rm cl}+\hbar\delta)
   \notag\\
   &=0.
\end{align}
This means that the following one is a chain complex. This is called \textit{quantum Batalin–Vilkovisky complex}.
\begin{align}
    {\rm Obs}^{\rm q}(U)= \Bigg(\cdots 
&\xrightarrow{\Delta_{\rm BV}^{\rm q}}\bigwedge^2 C_{\rm c}^\infty(U)^{-1}* {\rm Sym}\left(C_{\rm c}^\infty(U)^{0}\right)[\hbar]\notag\\
&\xrightarrow{\Delta_{\rm BV}^{\rm q}}C_{\rm c}^\infty(U)^{-1}* {\rm Sym}\left(C_{\rm c}^\infty(U)^{0}\right)[\hbar]\notag\\
&\xrightarrow{\Delta_{\rm BV}^{\rm q}}{\rm Sym}\left(C_{\rm c}^\infty(U)^{0}\right)[\hbar]\Bigg) 
\end{align}

${\rm Obs}^{\rm q}(U)$ does not have a Leibniz rule. Instead, we have the following theorem.

\begin{tcolorbox}[colframe=blue,colback=blue!3!]
\begin{theorem}
For $ A,B\in {\rm Obs}^{\rm q}(U)$ with some degree $|A|,|B|$,
\begin{align}
   \Delta_{\rm BV}^{\rm q}(A*B)=(\Delta_{\rm BV}^{\rm q}A)*B+(-1)^{|A|}A*(\Delta_{\rm BV}^{\rm q}B)+(-1)^{|A|}\hbar\{A,B\} .
\end{align}
\end{theorem}
\end{tcolorbox}
\noindent{}Here $\{-,-\}$ is the anti-bracket which is defined just below.\footnote{This definition is motivated by the usual anti-bracket in physics literature:
\begin{align}
 \{a,b\}:=\sum_{i=1}^N \left[\left(a\frac{\overleftarrow{\partial}}{\partial x^i}\right)\left(\frac{\overrightarrow{\partial}}{\partial {x^\star}^i}b\right)-\left(a\frac{\overleftarrow{\partial}}{\partial {x^\star}^i}\right)\left(\frac{\overrightarrow{\partial}}{\partial x^i}b\right)\right] \notag  
\end{align}
where $ x^i$ is bosonic and $ {x^\star}^i$ is fermionic.} 
By this theorem, $ H^*\left({\rm Obs}^{\rm q}(U)\right)$ does NOT have a product $*$ in contrast to $ H^*\left({\rm Obs}^{\rm cl}(U)\right)$. 
\begin{tcolorbox}[colframe=red,colback=red!3!]
\begin{definition}
        For $ f^{\star}\in C^{\infty}_{\rm c}(U)^{-1},\ g\in C^{\infty}_{\rm c}(U)^0$,
    we define anti-bracket:
    \begin{align}
      \{f^{\star},g\}:=\int_U f(x)g(x){\rm d}x\in\mathbb{R}\subset {\rm Sym}(C^{\infty}_{\rm c}(U)^0).
    \end{align}
    For any $A,B\in{\rm Obs}^{\rm cl,q}(U)$ with some definite degrees, we define anti-bracket which has the following properties:
    \begin{itemize}
        \item $\{A,B\}=0$ if $A,B\in C^{\infty}_{\rm c}(U)^0$ or $A,B\in C^{\infty}_{\rm c}(U)^{-1}$
        \item $\{A,B\}=-(-1)^{(|A|+1)(|B|+1)}\{B,A\}$
        \item $ \{A,B*C\}=\{A,B\}*C+(-1)^{(|A|+1)|B|}B*\{A,C\}$
    \end{itemize}
\end{definition}
\end{tcolorbox}

However in special situation we can define a product $*$ for quantum cohomology !
Let $A$ and $B$ have compact support in $U_1$ and $U_2$ respectively.
And assume $U_1$ and $U_2$ are disjoint open subsets of $U$. Then
\begin{align}
    \{A,B\}_U=0.\label{eq:anti_bracket_zero}
\end{align}
The subscription $U$ denotes that this is computed in $U$. 
This is because the computation of $\{ A,B\}_U$ reduces to the integrals
\begin{align}
    \int_{U}
    (U_1{\rm -supported\ function})
    \times
    (U_2{\rm -supported\ function})
    =0
\end{align}
(\ref{eq:anti_bracket_zero}) means that we have Leibniz rule for such $A,B$, thus we can define a product $*$ for them.

\begin{tcolorbox}[colframe=blue,colback=blue!3!]
\begin{theorem}\label{T0.3}
Let $U_1$ and $U_2$ be disjoint open subsets of $U$.
The quantum Batalin–Vilkovisky cohomology has a product. 
\begin{align}
\begin{array}{ccc}
H^*\left({\rm Obs}^{\rm q}(U_1)\right)\times H^*\left({\rm Obs}^{\rm q}(U_2)\right) & \stackrel{}{\longrightarrow} & H^*\left({\rm Obs}^{\rm q}(U)\right) \\
\rotatebox{90}{$\in$} & & \rotatebox{90}{$\in$} \\
([A]_{U_1},[B]_{U_2}) & \longmapsto & [A]_{U_1}*[B]_{U_2}=[A*B]_U
\end{array}
\end{align}
Here subscripts of $[\ \cdot\ ]_{-}$ denotes that it is in $H^*\left({\rm Obs}^{\rm q}(-)\right)$.
This product is the same as the ``operator product" in physics literature.
\end{theorem}
\end{tcolorbox}

The cohomology is formal series of $\hbar$, hence the definition of a state changes by a little.
\begin{tcolorbox}[colframe=red,colback=red!3!]
\begin{definition}
    A state $\langle-\rangle$ is a smooth map:
    \begin{align}
        \langle-\rangle:
        H^0\left({\rm Obs}^{\rm q}(M)\right)
        \to 
        \mathbb{C}[\hbar]
    \end{align}
\end{definition}
\end{tcolorbox}

\subsection{Weyl algebra in one-dimensional system}
In the case of $M=\mathbb{R}$, the quantum cohomology forms Weyl algebra (the canonical commutation relation).
\begin{tcolorbox}[colframe=blue,colback=blue!3!]
\begin{theorem}\label{thm:M=R_q}
    If $M=\mathbb{R}$, then for any value of $m$
    \begin{align}
        H^n\left({\rm Obs}^{\rm q}(\mathbb{R})\right)
        =
        \left\{
        \begin{array}{ll}
        {\rm Wey\ algebra} & (n=0) \\
        0 & ({\rm otherwise})
        \end{array}
        \right.
    \end{align}
\end{theorem}
\end{tcolorbox}
\noindent{}
The aim of this section is to explain what the generators of Weyl algebra are.
Roughly speaking, the generators of Weyl algebra are $\hbar,[Q],[P]$.\footnote{The definition of $Q$ and $P$ is written in section \ref{subsec:Q_P}.}
However, it seems strange, because  $Q,P$ have the same support $I=(-1/2,1/2)$, thus we can not define the products of them.

In order to take the products, we will define \textit{modified position observable} $\mathcal{Q}_t$ and \textit{modified momentum observable} $\mathcal{P}_t$.
\begin{tcolorbox}[colframe=red,colback=red!3!]
\begin{definition}
    Modified position and momentum  observable $\mathcal{Q}_t$ and $\mathcal{P}_t$ are in $C_{\rm c}^\infty(I_t),\ I_t=(-1/2+t,1/2+t)$. The definition of them is
\begin{align}
  \mathcal{Q}_t(x):=\phi_p(t)P_t(x)-\dot\phi_p(t)Q_t(x),
  \notag\\
  \mathcal{P}_t(x):=\phi_q(t)P_t(x)-\dot\phi_q(t)Q_t(x)
\end{align}
where $\phi_q$ and $\phi_p$ are defined in Appendix, and $Q_t$ and $P_t$ are defined as
\begin{align}
    Q_t(x):=Q(x-t),\ P_t(x):=P(x-t).
\end{align}
\end{definition}
\end{tcolorbox}
Modified observables have the following special property: 
\begin{tcolorbox}[colframe=blue,colback=blue!3!]
\begin{theorem}\label{thm:mathcal_Q_P_conserved}
\begin{align}
    \frac{\partial}{\partial t}
    \mathcal{Q}_t(\cdot)
    &=
    \Delta_{\rm BV}^{\rm q}
    \left(\phi_p(t)Q^\star_t(\cdot)\right),\\
    \frac{\partial}{\partial t}
    \mathcal{P}_t(\cdot)
    &=
    \Delta_{\rm BV}^{\rm q}
    \left(\phi_q(t)Q^\star_t(\cdot)\right).
\end{align}
Remind that $Q=\mathcal{Q}_{t=0}$ and $P=\mathcal{P}_{t=0}$, then for any $t$
\begin{align}
    [Q]=[\mathcal{Q}_t],\ \ 
    [P]=[\mathcal{P}_t].
\end{align}
\end{theorem}
\end{tcolorbox}
\begin{proof}
We compute $\frac{\partial}{\partial t}\mathcal{Q}_t(x)$.
\begin{align}
    \frac{\partial}{\partial t}\mathcal{Q}_t(x) 
    &=\dot\phi_p(t)P_t(x)+\phi_p(t)\frac{\partial}{\partial t}P_t(x)-\ddot\phi_p(t)Q_t(x)-\dot\phi_p(t)\frac{\partial}{\partial t}Q_t(x) \notag\\
    &=-\dot\phi_p(t)\frac{\partial}{\partial x}Q(x-t)-\phi_p(t)\frac{\partial}{\partial t}\frac{\partial}{\partial x}Q(x-t)-\ddot\phi_p(t)Q(x-t)-\dot\phi_p(t)\frac{\partial}{\partial t}Q(x-t) \notag\\
    &=-\phi_p(t)\frac{\partial}{\partial t}\frac{\partial}{\partial x}Q(x-t)-\ddot\phi_p(t)Q(x-t) \notag\\
    &=-\phi_p(t)\frac{\partial}{\partial t}\frac{\partial}{\partial x}Q(x-t)-m^2\phi_p(t)Q(x-t) \notag\\ &=\phi_p(t)\left(\frac{\partial}{\partial x}\right)^2Q(x-t)-m^2\phi_p(t)Q(x-t) \notag\\
    &=\left[\left(\frac{\partial}{\partial x}\right)^2-m^2\right]\phi_p(t)Q(x-t) \notag\\
    &=\Delta_{\rm BV}^{\rm q}\left(\phi_p(t)Q^\star_t(x)\right).
\end{align}
By the similar calculations, we also obtain
\begin{align}
    \frac{\partial}{\partial t}\mathcal{P}_t(x)
    =
    \Delta_{\rm BV}^{\rm q}\left(\phi_q(t)Q^\star_t(x)\right).
\end{align}
\end{proof}

The support of $\mathcal{Q}_t$ and $\mathcal{P}_t$ is $I_t=(-1/2+t,1/2+t)$. 
Therefore we can define products of them like
\begin{align}
    [\mathcal{Q}_t]_{I_t}*[\mathcal{P}_s]_{I_s}
    \ \ 
    {\rm or}
    \ \ 
    [\mathcal{Q}_t*\mathcal{P}_s]
    \ \ 
    (|t-s|>1)
\end{align}
where $I_t$ and $I_s$ denote being in $H^*({\rm Obs}^{\rm q}(I_t))$ and $H^*({\rm Obs}^{\rm q}(I_s))$.
Note that this product is independent for the choice $t$ and $s$ because of Theorem \ref{thm:mathcal_Q_P_conserved}.

\begin{tcolorbox}[colframe=blue,colback=blue!3!]
\begin{theorem}
    We have the canonical commutation relation:
    \begin{align}
        [\mathcal{P}_t*\mathcal{Q}_s]
        -
        [\mathcal{Q}_t*\mathcal{P}_s]
        =\hbar[1]
        \ \ 
         (s-t>1).
    \end{align}
    $[\mathcal{Q}_\bullet]$ and $[\mathcal{P}_\bullet]$ are the generator. Then  $H^0({\rm Obs}^{\rm q}(\mathbb{R}))$ is Weyl algebra.
\end{theorem}
\end{tcolorbox}
\begin{proof}
    It is enough to show that for some $t>1$
    \begin{align}
      [Q_0*(\mathcal{P}_{-t}-\mathcal{P}_{t})]
      =
      \hbar[1].  
    \end{align}
    Reminding that
    \begin{align}
        \frac{\partial}{\partial t}\mathcal{P}_t(x)
        =
        \Delta_{\rm BV}^{\rm q}\left(\phi_q(t)Q^\star_t(x)\right)
    \end{align}
    and integrating over $[t_1,t_2]$, then we have
    \begin{align}
        \mathcal{P}_{t_2}-\mathcal{P}_{t_1}
        &=
        \Delta_{\rm BV}^{\rm q}h_{t_1,t_2}^\star ,\\
 h_{t_1,t_2}^\star(x)&:=\int_{t_1}^{t_2}{\rm d}t\ \phi_q(t)Q^\star_t(x)\in C_{\rm c}^\infty(\mathbb{R})^{-1}.
    \end{align}
    We have defined $ h_{t_1,t_2}(x):=\int_{t_1}^{t_2}{\rm d}t\ \phi_q(t)Q_t(x)\in C_{\rm c}^\infty(\mathbb{R})^{0}$.

Let us think of 
\begin{align}
   S_t:=Q_0*h_{-t,t}^\star\in C_{\rm c}^\infty(\mathbb{R})^{0}*C_{\rm c}^\infty(\mathbb{R})^{-1}. 
\end{align}
$\Delta_{\rm BV}^{\rm q}$ acting on it,
\begin{align}
    \Delta_{\rm BV}^{\rm q}S_t 
 &=
 Q_0*(\Delta_{\rm BV}^{\rm q}h_{-t,t}^\star)+\hbar\int_{\mathbb{R}}{\rm d}x\ Q_0(x)h_{-t,t}(x) \notag\\
 &=
 Q_0*(\mathcal{P}_{t}-\mathcal{P}_{-t})+\hbar\int_{\mathbb{R}}{\rm d}x\int_{-t}^{t}{\rm d}u\ Q_0(x)\phi_q(u)\mathcal{Q}_{u}(x) \notag\\
 &=
 Q_0*(\mathcal{P}_{t}-\mathcal{P}_{-t})+\hbar\int_{\mathbb{R}}{\rm d}x\int_{-t}^{t}{\rm d}u\ Q(x)\phi_q(u)Q(x-u).
\end{align}
Therefore we will show that
\begin{align}
    \hbar\int_{\mathbb{R}}{\rm d}x\int_{-t}^{t}{\rm d}u\ Q(x)\phi_q(u)Q(x-u)=\hbar
\end{align}
for some $t>1$.

Now we take a limit $t\to\infty$\footnote{When $u$ is sufficiently large, $Q(x)Q(x-u)=0$, then the integrand goes to zero. Therefore we can take the limit $t\to \infty$.}.
\begin{align}
\Delta_{\rm BV}^{\rm q}S_\infty  
=
Q_0*(\mathcal{P}_{\infty}-\mathcal{P}_{-\infty})+\hbar\int_{\mathbb{R}}{\rm d}x\int_{\mathbb{R}}{\rm d}u\ Q(x)\phi_q(u)Q(x-u) 
\end{align}
Since $Q$ is an even function,
\begin{align}
\Delta_{\rm BV}^{\rm q}S_\infty  
=
Q_0*(\mathcal{P}_{\infty}-\mathcal{P}_{-\infty})+\hbar\int_{\mathbb{R}}{\rm d}x\int_{\mathbb{R}}{\rm d}u\ Q(x)\phi_q(u)Q(u-x). 
\end{align}
By a change of variable: $u\to u+x$
\begin{align}
\Delta_{\rm BV}^{\rm q}S_\infty  
=
Q_0*(\mathcal{P}_{\infty}-\mathcal{P}_{-\infty})+\hbar\int_{\mathbb{R}}{\rm d}x\int_{\mathbb{R}}{\rm d}u\ Q(x)\phi_q(u+x)Q(u)\label{eq:can_rel_proof_1}
\end{align}
By the assumption, for $m\neq 0$,
\begin{align}
   \int_{\mathbb{R}}{\rm d}x\ Q(x)\phi_q(x)=\frac{1}{2}\int_{\mathbb{R}}{\rm d}x\ Q(x)(e^{mx}+e^{-mx})=1 
\end{align}
and $Q$ is even, hence
\begin{align}
   \int_{\mathbb{R}}{\rm d}x\ Q(x)e^{mx}=\int_{\mathbb{R}}{\rm d}x\ Q(x)e^{-mx}=1.\label{eq:can_rel_proof_2}
\end{align}
The second term of the right hand side of (\ref{eq:can_rel_proof_1}),
\begin{align}
   \hbar\int_{\mathbb{R}}{\rm d}x\int_{\mathbb{R}}{\rm d}u\ Q(x)\phi_q(u+x)Q(u) 
&=\frac{\hbar}{2}\int_{\mathbb{R}}{\rm d}x\int_{\mathbb{R}}{\rm d}u\ Q(x)(e^{m(x+u)}+e^{-m(x+u)})Q(u) \notag\\
&=\hbar .
\end{align}
We used (\ref{eq:can_rel_proof_2}) for the last line.

In the case of $m=0$, since $\phi_q=1$, we can easily check that the second term of right hand side of (\ref{eq:can_rel_proof_1}) is $\hbar$.
\end{proof}

\section{Construction of topological operator}
\subsection{Shift symmetry}
In case of $m=0$, we have a shift symmetry
\begin{align}
    \Phi\mapsto\Phi+\alpha\ \ 
    (\alpha\in\mathbb{R})
\end{align}
for equation of motion $\Delta\Phi=0$.
Reflecting it, $P_t$ must be conserved.
This is true because $\mathcal{P}_t$ is conserved and in massless case
\begin{align}
    \mathcal{P}_t(x)
    =
    \phi_q(t)P_t(x)-\dot\phi_q(t)Q_t(x)
    =
    P_t(x)
\end{align}
In other words, $P_t$ is the Noether charge.
\subsection{Construction of topological operator}

First of all, we will review the usual construction of the topological operators.
Let $\hat{q}$ and $\hat{p}$ be the generators of Wely algebra, i.e., satisfy $[\hat{q},\hat{p}]=\hbar$.
In physics literature, the topological operator of the shift symmetry is
\footnote{We have no $i$ in front of the $\alpha \hat{p}$, because Weyl algebra is $[\hat{q},\hat{p}]=\hbar$.}
\begin{align}
    \hat{V}_\alpha
    :=\exp(\alpha \hat{p})=\sum_{n=0}^\infty \frac{1}{n!}(\alpha\hat{p})^n.
\end{align}
Obviously we have
\begin{align}
     \hat{q}\hat{V}_{\alpha}
     =\hat{V}_{\alpha}(\hat{q}+\alpha\hbar)
     \because\ 
     \hat{q}\hat{p}^n
     =
     \hat{p}^n\hat{q}
     +
     n\hbar\hat{p}^{n-1}.
     \label{eq:hat_q_V}
\end{align}

Then the naive construction of the topological operator in Batalin-Vilkovisky formalism is
\begin{align}
    [\mathcal{V}_{\alpha,t}]
    &:=
    \sum_{n=0}^\infty \frac{1}{n!}\ \alpha^n
    [P_{s_0+t}*P_{s_1+t}*\cdots*P_{s_n+t}],\notag\\ 
    &(s_1-s_0>1,\ s_2-s_1>1,\cdots,s_n-s_{n-1}>1).
\end{align}
The supports of $I_{s_0+t},I_{s_0+t},\cdots$ are disjoint.
Hence each term of $[\mathcal{V}_{\alpha,t}]$ works well in terms of ``topologicalness" and ``action for $Q_0$".
\begin{itemize}
    \item Topologicalness\\ 
    By using Theorem \ref{thm:mathcal_Q_P_conserved} and $\mathcal{P}_t=P_t$, 
    \begin{align}
        \frac{{\rm d}}{{\rm d}t}
        [P_{s_0+t}*P_{s_1+t}*\cdots*P_{s_n+t}]=0.
    \end{align} 
    \item Action for $Q_0$ \\
    Similar to (\ref{eq:hat_q_V}), we have
    \begin{align}
        \left[
        Q_0*
        P_{s_0+t_+}*P_{s_1+t_+}*\cdots*P_{s_n+t_+}
        \right]
        =&
        \left[
        P_{s_0+t_-}*P_{s_1+t_-}*\cdots*P_{s_n+t_-}
        *Q_0
        \right]
        \notag\\
        &+
        n\hbar
        \left[
         P_{s_0+t_-}*P_{s_1+t_-}*\cdots*P_{s_{n-1}+t_-}
        \right]
    \end{align}
    where $t_+$ is sufficiently large and $t_-$ is sufficiently small.
\end{itemize}
However there is a problem.
The support of $\mathcal{V}_\alpha$ is infinitely wide.
This is not good as the topological operator because we can not define the action.

Therefore we define the topological operator in another way.
\footnote{${\rm Obs}^{\rm q}(U)$ is made of ${\rm Sym}(C_{\rm c}^\infty(U))$. If we take ``${\rm Sym}$" literally, ${\rm Obs}^{\rm q}(U)$ has only polynomials. Hence $\mathcal{U}_{\alpha,t}$ is technically outside of the definition of ${\rm Obs}^{\rm q}(U)$. However we take a completion implicitly \ref{ref:completion}. It might make $\mathcal{U}_{\alpha,t}$ is ${\rm Obs}^{\rm q}(U)$.}
\begin{tcolorbox}[colframe=red,colback=red!3!]
\begin{definition}
\begin{align}
    [\mathcal{U}_{\alpha,t}]
    :=
    \sum_{n=0}^\infty \frac{1}{n!}\ \alpha^n
    [\underbrace{P_t*P_t*\cdots*P_t}_{n}]
    =
    \sum_{n=0}^\infty \frac{1}{n!}\ \alpha^n
    [P_t^{*n}]
\end{align}
$\mathcal{U}_{\alpha,t}$ obviously has the finite support $I_t$.    
\end{definition}
\end{tcolorbox}

Instead of the finite support, we lost the clearness of the following properties.
\begin{itemize}
    \item Topologicalness 
    \begin{align}
        \left[\frac{{\rm d}}{{\rm d}t}\mathcal{U}_{\alpha,t}\right]_{I_t}=0
    \end{align} 
    \item Action for $Q_0$ 
    \begin{align}
        [Q_0*\mathcal{U}_{\alpha,t}]
        =
        [\mathcal{U}_{\alpha,s}*(Q_0+\alpha\hbar)],\ \  (t>1,-1>s)
    \end{align}
\end{itemize}
In the next sections, we will prove these properties.

\subsection{Proof of topologicalness}
\begin{tcolorbox}[colframe=blue,colback=blue!3!]
\begin{theorem}\label{Thm:topological_operator}
    We have a topological operator $\mathcal{U}_{\alpha,t}$. Thus
    \begin{align}
        \left[\frac{{\rm d}}{{\rm d}t}\mathcal{U}_{\alpha,t}\right]_{I_t}=0.
    \end{align} 
\end{theorem}    
\end{tcolorbox}
\begin{proof}
    It is enough to show that
    \begin{align}
        \frac{{\rm d}}{{\rm d}t}[P_t^{*n}]_{I_t}=0.
    \end{align}   
    We have
    \begin{align}
       &\frac{{\rm d}}{{\rm d}t}P_t^{*n}
       =
       nP_t^{*(n-1)}*\frac{{\rm d}}{{\rm d}t}P_t,\\ 
       &\frac{{\rm d}}{{\rm d}t}P_t
       =
       \Delta_{\rm BV}^{\rm q}\left(\phi_q(t)Q^\star_t\right)
       =
       \Delta_{\rm BV}^{\rm q}Q^\star_t
    \end{align}
    since $\phi_q=1$ for the case of $m=0$.
    Then we obtain
    \begin{align}
      \frac{{\rm d}}{{\rm d}t}P_t^{*n}=nP^{*(n-1)}_t*\Delta_{\rm BV}^{\rm q}Q^\star_t . 
    \end{align}   
    It is enough to show that
    \begin{align}
        \Delta_{\rm BV}^{\rm q}(P_t^{*(n-1)}*Q_t^\star)=P_t^{*(n-1)}*\Delta_{\rm BV}^{\rm q}Q_t^\star.
    \end{align}
    However in fact $P_t^{*(n-1)}\in {\rm Sym}(C^\infty_{\rm c}(I_t)^0),\ Q^\star_t\in C_{\rm c}^\infty(I_t)^{-1}$ then
    \begin{align}
        \Delta_{\rm BV}^{\rm q}(P_t^{*(n-1)}*Q_t^\star)=P_t^{*(n-1)}*\Delta_{\rm BV}^{\rm q}Q_t^\star+\hbar\delta(P_t^{*(n-1)}*Q_t^\star).
    \end{align}
    Hence we need to see that $\delta(P_t^{*(n-1)}*Q_t^\star)$ vanishes.
    \begin{align}
        \delta(P_t^{*(n-1)}*Q_t^\star)
        =(n-1) P_t^{*(n-2)}\int_{\mathbb{R}}P_t(x)Q_t(x){\rm d}x
    \end{align}
    We will see that$\int_{\mathbb{R}}P_t(x)Q_t(x){\rm d}x$ vanishes.
    \begin{align}
      \int_{\mathbb{R}}P_t(x)Q_t(x){\rm d}x
      &=-\int_{\mathbb{R}}Q'(x-t)Q(x-t){\rm d}x\notag\\
      &=-\int_{\mathbb{R}}Q'(x)Q(x){\rm d}x\notag\\
      &=-\frac{1}{2}\int_{\mathbb{R}}
      \frac{\partial}{\partial x}(Q(x)Q(x)){\rm d}x
    \end{align}
    $Q(x)$ has compact support in $I$, so $\int_{\mathbb{R}}P_t(x)Q_t(x){\rm d}x=0.$
\end{proof}

\subsection{Proof of action for operators}
\begin{tcolorbox}[colframe=blue,colback=blue!3!]
\begin{theorem}\label{thm:Q_0_charged}
    $Q_0$ is a charged operator against $\mathcal{U}_{\alpha,t}$.
    \begin{align}
        [Q_0*\mathcal{U}_{\alpha,t}]
        =
        [\mathcal{U}_{\alpha,s}*(Q_0+\alpha\hbar)],
        \ \  (t>1,-1>s)
    \end{align}
\end{theorem}
\end{tcolorbox}
\begin{proof}
    It is enough to show that
    \begin{align}
       [Q_0*(P_{t}^{*n}-P_{-t}^{*n})]=n\hbar[P_0^{*(n-1)}] 
    \end{align}
    for some $t>1$.
    We have obtained
    \begin{align}
        \frac{{\rm d}}{{\rm d}t}P_t^{*n}=n\Delta_{\rm BV}^{\rm q}(P_t^{*(n-1)}*Q_t^\star) 
    \end{align}
    in the proof of Theorem \ref{Thm:topological_operator}.
    Hence we have
    \begin{align}
    &P_{t_2}^{*n}-P_{t_1}^{*n}=\Delta_{\rm BV}^{\rm q}H_{t_1,t_2}(n), \notag\\
    &H_{t_1,t_2}(n):=n\int_{t_1}^{t_2}{\rm d}u\ P_u^{*(n-1)}*Q_u^\star. 
    \end{align}
    By multiplication of $Q_0$
    \begin{align}
        Q_0*(P_{t}^{*n}-P_{-t}^{*n})=Q_0*\Delta_{\rm BV}^{\rm q}H_{-t,t}(n).
    \end{align}
    We will show that the right hand side is $n\hbar P_0^{*(n-1)}$ up to the cohomology.

    We know $\delta(P_u^{*(n-1)}*Q_u^\star)=0$ from the proof of Theorem \ref{thm:Q_0_charged}, 
    then we have
    \begin{align}
        \Delta_{\rm BV}^{\rm q}(Q_0*H_{-t,t}(n)) 
        =Q_0*\Delta_{\rm BV}^{\rm q}(H_{-t,t}(n))+\hbar\delta(Q_0*H_{-t,t}(n)), 
    \end{align}
    thus it is enough to show that $\hbar\delta(Q_0*H_{-t,t})$ is equivalent to $n\hbar P_0^{*(n-1)}$ up to the cohomology.
    
    We compute $\hbar\delta(Q_0*H_{-t,t}(n))$.
    \begin{align}
        \hbar\delta(Q_0*H_{-t,t}(n))
        &=
        n\hbar\int_{-t}^{t}{\rm d}u\ \delta(Q_0*P_u^{*(n-1)}*Q_u^\star)\notag\\
        &=
        n\hbar\int_{-t}^{t}{\rm d}u\ 
        \left(
        \delta(Q_0*Q_u^\star)P_u^{*(n-1)}
        +
        Q_0*\delta(P_u^{*(n-1)}*Q_u^\star)
        \right)\notag\\
        &=
        n\hbar\int_{-t}^{t}{\rm d}u\ 
        \delta(Q_0*Q_u^\star)P_u^{*(n-1)} \notag\\
        &=
        n\hbar\int_{-t}^{t}{\rm d}u\ 
        P_u^{*(n-1)}\int_{\mathbb{R}}{\rm d}x Q_0(x)Q_u(x) \notag\\
        &=
        n\hbar\int_{-t}^{t}{\rm d}u\ 
        P_u^{*(n-1)}\int_{\mathbb{R}}{\rm d}x Q(x)Q(x-u) 
    \end{align}
    Here we used $\delta(P_u^{*(n-1)}*Q_u^\star)=0$.
    Substituting $P_{u}^{*(n-1)}-P_{c}^{*(n-1)}=\Delta_{\rm BV}^{\rm q}\tilde{H}_{c,u}(n-1)$,
    \begin{align}
        \hbar\delta(Q_0*H_{-t,t}(n))
        &=
        n\hbar P_c^{*(n-1)}\int_{-t}^{t}{\rm d}u\ 
        \int_{\mathbb{R}}{\rm d}x Q(x)Q(x-u)
        \notag\\
        &+
        n\hbar\int_{-t}^{t}{\rm d}u\ 
        \Delta_{\rm BV}^{\rm q}\tilde{H}_{c,u}(n-1)\int_{\mathbb{R}}{\rm d}x Q(x)Q(x-u)
    \end{align}
    In the second term of the right hand side, $Q(x)Q(x-u)$ is just a number\footnote{These trivially act for the scalar field $\Phi\in C^\infty(\mathbb{R})$}, thus this term is trivial in cohomology.
    It remains to show that
    \begin{align}
        \int_{-t}^{t}{\rm d}u\ 
        \int_{\mathbb{R}}{\rm d}x Q(x)Q(x-u)=1.
    \end{align}
    for some $t>1$.
    When $u$ is sufficiently large, $Q(x)Q(x-u)\to 0$.
    Therefore we can take a limit $t\to\infty$.
    \begin{align}
        \int_{\mathbb{R}}{\rm d}u\ 
        \int_{\mathbb{R}}{\rm d}x Q(x)Q(x-u)
        =
        \int_{\mathbb{R}}{\rm d}u\ 
        \int_{\mathbb{R}}{\rm d}x Q(x)Q(u-x)
        =
        \int_{\mathbb{R}}{\rm d}u\ 
        \int_{\mathbb{R}}{\rm d}x Q(x)Q(u)
        =1.
    \end{align}
\end{proof}
\begin{tcolorbox}[colframe=blue,colback=blue!3!]
\begin{theorem}
    $P_0$ has no charge against $\mathcal{U}_{\alpha,t}$.
    \begin{align}
        [P_0*\mathcal{U}_{\alpha,t}]
        =
        [\mathcal{U}_{\alpha,s}*P_0],
        \ \  (t>1,-1>s)
    \end{align}
\end{theorem}
\end{tcolorbox}
\begin{proof}
    It is enough to show that
    \begin{align}
       [P_0* (P^{*n}_{t}-P_{-t}^{*n})]=0 
    \end{align}
    for some $t>1$. We already have
\begin{align}
    &P_{t_2}^{*n}-P_{t_1}^{*n}=\Delta_{\rm BV}^{\rm q}H_{t_1,t_2},\\
   &H_{t_1,t_2}:=n\int_{t_1}^{t_2}{\rm d}u\ P_u^{*(n-1)}*Q_u^\star. 
\end{align}
By multiplication of $P_0$
\begin{align}
    P_0*(P_{t_2}^{*n}-P_{t_1}^{*n})=P_0*\Delta_{\rm BV}^{\rm q}H_{t_1,t_2}.
\end{align}
We will show the right hand side is zero up to cohomology.

Since
\begin{align}
   \Delta_{\rm BV}^{\rm q}(P_0*H_{t_1,t_2})
   =
   P_0*\Delta_{\rm BV}^{\rm q}(H_{t_1,t_2})
   +
   \hbar\delta(P_0*H_{t_1,t_2}),
\end{align}
it is enough to show that $\hbar\delta(P_0*H_{t_1,t_2})=0$ up to the cohomology.

We compute $\hbar\delta(P_0*H_{t_1,t_2})$.
\begin{align}
    \hbar\delta(P_0*H_{t_1,t_2})
    &=
    \hbar\int_{-t}^{t}{\rm d}u\ \delta(P_0*P_u^{*(n-1)}*Q_u^\star)\notag\\
    &=
    \hbar\int_{-t}^{t}{\rm d}u\ 
    \left(
    \delta(P_0*Q_u^\star)P_u^{*(n-1)}
    +
    P_0*\delta(P_u^{*(n-1)}*Q_u^\star)
    \right)\notag\\
    &=
    \hbar\int_{-t}^{t}{\rm d}u\ \delta(P_0*Q_u^\star)P_u^{*(n-1)}\notag\\
    &=
    \hbar\int_{-t}^{t}{\rm d}u\ 
    P_u^{*(n-1)}\int_{\mathbb{R}}{\rm d}x\ P_0(x)Q_u(x)
    \notag\\
    &=
    \hbar\int_{-t}^{t}{\rm d}u\ 
    P_u^{*(n-1)}\int_{\mathbb{R}}{\rm d}x\ Q'(x)Q(x-u)
\end{align}
Here we used $\delta(P_u^{*(n-1)}*Q_u^\star)=0$. 
Substituting $P_{u}^{*(n-1)}-P_{c}^{*(n-1)}=\Delta_{\rm BV}^{\rm q}H_{c,u}$,
\begin{align}
    \hbar\delta(P_0*H_{t_1,t_2})
    &=
    P_c^{*(n-1)}
    \hbar\int_{-t}^{t}{\rm d}u\ 
    \int_{\mathbb{R}}{\rm d}x\ Q'(x)Q(x-u)\notag\\
    &+
    \hbar\int_{-t}^{t}{\rm d}u\ 
    (\Delta_{\rm BV}^{\rm q}H_{c,u})\int_{\mathbb{R}}{\rm d}x\ Q'(x)Q(x-u)
\end{align}
    In the second term of the right hand side, $Q'(x)Q(x-u)$ is just a number, thus this term is trivial in cohomology.
    It remains to show that
    \begin{align}
        \int_{-t}^{t}{\rm d}u\ 
        \int_{\mathbb{R}}{\rm d}x Q'(x)Q(x-u)=0.
    \end{align}
    for some $t>1$.
    When $u$ is sufficiently large, $Q'(x)Q(x-u)\to 0$.
    Hence we can take a limit $t\to\infty$.
    \begin{align}
        \int_{\mathbb{R}}{\rm d}u\ 
        \int_{\mathbb{R}}{\rm d}x Q'(x)Q(x-u)
        =
        \int_{\mathbb{R}}{\rm d}u\ 
        \int_{\mathbb{R}}{\rm d}x Q'(x)Q(u-x)
        =
        \int_{\mathbb{R}}{\rm d}u\ 
        \int_{\mathbb{R}}{\rm d}x Q'(x)Q(u)
        =0.
    \end{align}
\end{proof}

\section{Gauging and compact scalar}
In this section, we will give an application of the topological operator.
Some discussion is not established rigorously, but physically natural.

\subsection{$\mathbb{Z}$-gauging}
Topological operators have the advantage over Noether charge, thus we can define \textit{discrete gauging}.\footnote{More precisely, we need to show the fusion rule : $\mathcal{U}_{\alpha}\mathcal{U}_{\beta}=\mathcal{U}_{\alpha+\beta}$ to make the discussion of this section complete.} This is because topological operators describe finite group transformations, while Noether charges only do infinitesimal transformations. 
Hence now we have group action like
\begin{align}
    \mathbb{R}\curvearrowright
    H^*({\rm Obs}^{\rm q}(\mathbb{R})).
\end{align}
Subgroup $\mathbb{Z}\subset \mathbb{R}$ also acts as
\begin{align}
    \mathbb{Z}\curvearrowright
    H^*({\rm Obs}^{\rm q}(\mathbb{R})).
\end{align}
Hence we can define\footnote{As (pre)factorization algebra, it is better to consider finite interval: $I\leadsto H^*({\rm Obs}^{\rm q}(I))/\mathbb{Z}$.}
\begin{align}
    \frac
    {H^*({\rm Obs}^{\rm q}(\mathbb{R}))}
    {\mathbb{Z}}
\end{align}
The generators are
\begin{align}
    \hbar,\ \ 
    \left[{\rm Exp}(\pm iQ)\right],\ \ 
    [P]
\end{align}
where ${\rm Exp}(\bullet):=\sum_n\frac{1}{n!}\bullet ^{*n}$. 
They satisfy
\begin{align}
    [{\rm Exp}(\pm iQ)*P_{-t}]
    -
    [{\rm Exp}(\pm iQ)*P_{t}]
    =[{\rm Exp}(\pm iQ)]\ \ (t>1).
\end{align}
This is the same as $[\hat{p},e^{\pm i\hat{q}}]=\hbar e^{\pm i\hat{q}}$ in a usual notation.
\begin{tcolorbox}[colframe=red,colback=red!3!]
\begin{definition}
We saw the algebra generated by
\begin{align}
    \hat{p},e^{i\hat{q}},e^{-i\hat{q}},\hbar 1
\end{align}
and satisfying
\begin{align}
    [\hat{p},e^{\pm i\hat{q}}]=\hbar e^{\pm i\hat{q}}.
\end{align}
We call it periodic Weyl algebra.
\end{definition}
\end{tcolorbox}
We can say
\begin{align}
    \frac
    {H^0({\rm Obs}^{\rm q}(\mathbb{R}))}
    {\mathbb{Z}}
    \cong
    {\rm periodic\ Weyl\ algebra}
\end{align}
similarly to Theorem \ref{thm:M=R_q}.

\subsection{Compact scalar and $\theta$-vacuum}
After $\mathbb{Z}$ gauging, massless scalar theory becomes compact scalar theory as explained in \cite{Fukushima:2022zor},\cite{Gorantla:2021svj}.
In compact scalar theory, we have $\theta$-degree. 
We can express $\theta$ as $\theta$-term or $\theta$-vacuum.
However to implement $\theta$-term is difficult in this formalism, because it is based on equation of motion and equation of motion does not have $\theta$-term.
On the other hand, implementing $\theta$-vacuum is relatively easy since a vacuum is just a map from the cohomology to $\mathbb{C}[\hbar]$.
\begin{align}
    \frac
    {H^*({\rm Obs}^{\rm q}(\mathbb{R}))}
    {\mathbb{Z}}
    \to
    \mathbb{C}[\hbar]\label{eq:map_Z-gauged_to_C-hbar}
\end{align}

One way to give a state is to think a representation of $H^*({\rm Obs}^{\rm q}(\mathbb{R}))/\mathbb{Z}$.
For simplicity, we use $e^{\pm i\hat{q}}$ and $\hat{p}$ rather than $[{\rm Exp}(\pm iQ)]$ and $[P]$.
Take eigenvector $v_p$ 
\begin{align}
    \hat{p}v_p=pv_p.
\end{align}
Then we find that $p$ must be quantized.
To see that, consider a vector $e^{i\hat{q}}v_p$. This satisfies
\begin{align}
    \hat{p}(e^{i\hat{q}}v_p)
    =
    (p+\hbar)(e^{i\hat{q}}v_p)
\end{align}
Therefore $e^{\pm i\hat{q}}$ are ladder operators.
The spectrum of $\hat{p}$ is
\begin{align}
    \lambda_r(\hat{p})=\{\hbar(n+r)\ |\ n\in \mathbb{Z},\  r\in \mathbb{R}\}
\end{align}.
which depend on $r$. Obviously $r+1$-spectrum is equivalent to  $r$-spectrum.
Physicists denote $r$ as $\theta/2\pi$.
At $\theta=0$ and $\theta=\pi$, we can see $\mathbb{Z}_2$-symmetry of the spectrum : $p\mapsto-p$.
Hence the map ($\ref{eq:map_Z-gauged_to_C-hbar}$) is given by a trace
\begin{align}
    \langle-\rangle_{r=\frac{\theta}{2\pi}}:X\mapsto\sum_{p\in \lambda_r(\hat{p})} (v_p,Xv_p).
\end{align}
We can see $\theta$-dependence in the state.

\section{Conclusion and discussion}
We have seen the construction of a topological operator $\mathcal{U}_{\alpha,t}$.
The point of the construction is the order: at first taking products $*$, and next taking a cohomology $[\cdot]$ (we call it the order A). 
If we take reversed order (we call it the order B), we have
\begin{align}
    [\mathcal{V}_\alpha]
    &:=
    \sum_{n=0}^\infty \frac{1}{n!}(i\alpha)^n
    [P]_{I_{s_0}}*[P]_{I_{s_1}}*\cdots*[P]_{I_{s_n}},\notag\\ 
    &(s_1-s_0>1,\ s_2-s_1>1,\cdots,s_n-s_{n-1}>1),
\end{align}
and it has infinitely wide support.
The order A vs the order B is the similar to that path integral formalism vs operator formalism.
In path integral formalism, we can formally think 
of multiplications of operators like $\mathcal{O}_1(\Phi)\mathcal{O}_2(\Phi)\cdots\mathcal{O}_n(\Phi)$ because these are just c-numbers.
On the other hand, in operator formalism, in order to give multiplications of operators, we need to get rid of UV divergences like $:\mathcal{O}_1\mathcal{O}_2\cdots\mathcal{O}_n:$ since they are q-numbers.
In the order A, observables $\mathcal{O}$ are just c-number, thus we can freely take the products $\mathcal{O}_1*\mathcal{O}_2*\cdots*\mathcal{O}_n$. This is the same as path integral formalism. However, in the order B, $[\mathcal{O}]$ can not be taken products freely, therefore this corresponds to operator formalism.
In order to take products, the support of $[\mathcal{O}_1]*[\mathcal{O}_2]*\cdots*[\mathcal{O}_n]$ need to be wider than the original support. It is a kind of UV divergences.

And we have seen $\theta$-dependence of the state $\langle-\rangle$ by representation theory.
It is interesting to construct map in another way.
For example, the state of the massive scalar theory on $M=\mathbb{R}$ can be given by the embedding
\begin{align}
    f:C_{\rm c}^\infty(\mathbb{R})
    \hookrightarrow
    \mathcal{S}(\mathbb{R})
\end{align}
where $\mathcal{S}(\mathbb{R})$ is the set of Schwartz functions on $\mathbb{R}$.
By considering ${\rm Obs}_{\mathcal{S}}:={\rm Sym}(\mathcal{S}(\mathbb{R})^{-1}\to \mathcal{S}(\mathbb{R})^{0})$, we obtain $H^0({\rm Obs}^{\rm q}_{\mathcal{S}}(\mathbb{R})))\cong\mathbb{C}[\hbar]$. Then $f$ induce the state
\begin{align}
    H^0({\rm Obs}^{\rm q}(\mathbb{R}))
    \to
    H^0({\rm Obs}^{\rm q}_{\mathcal{S}}(\mathbb{R}))\cong\mathbb{C}[\hbar]
\end{align}
It is attractive to give a state in similar way in the case of compact scalar.
However, in order to achieve this, we need the technique to remove IR divergence because compact scalar is massless. Such a technique is discussed in \cite{Kawahira:2024}.

And there is another way to consider the compact scalar theory. It is to focus on factorization homology.
The factorization algebra of compact scalar is essentially a functor which assigns periodic Weyl algebra to each open subsets.
Hence if we consider this system in $S^1$, the factorization homology is the same as Hochschild homology of periodic Weyl algebra. 
To extract the $\theta$-information from the Hochschild homology is a fascinating work.

\section*{Acknowledgment}
We would like to thank Yuji Ando, Hayato Kanno, Keisuke Konosu, Hiroaki Matsunaga and Jojiro Totsuka-Yoshinaka for the discussions at the early stage of this work.
We would also like to thank Jun Maeda and Yuto Moriwaki for helpful discussions.
The work of MK is supported by Grant-in-Aid for JSPS Fellows No. 22KJ1989.


\bibliographystyle{ptephy}
\bibliography{sample}

\begin{thebibliography}{10}

\bibitem{Costello:2016vjw}
Kevin Costello and Owen Gwilliam,
\newblock {\em {Factorization Algebras in Quantum Field Theory}},
\newblock  (Cambridge University Press, 12 2016).

\bibitem{Costello:2021jvx}
Kevin Costello and Owen Gwilliam,
\newblock {\em {Factorization Algebras in Quantum Field Theory}},
\newblock New Mathematical Monographs (41).  (Cambridge University Press, 9 2021).

\bibitem{Alfonsi:2023qpv}
Luigi Alfonsi and Charles A.~S. Young (7 2023),  {{arXiv:2307.15106}}.

\bibitem{Elliott:2014haa}
Chris Elliott, Math. Phys. Anal. Geom., {\bf 22}(4), 22 (2019),  {{arXiv:1402.0890}}.

\bibitem{Costello:2023knl}
Kevin Costello and Owen Gwilliam (10 2023),  {{arXiv:2310.06137}}.

\bibitem{Kapustin:2014gua}
Anton Kapustin and Nathan Seiberg, JHEP, {\bf 04}, 001 (2014),  {{arXiv:1401.0740}}.

\bibitem{Gaiotto:2014kfa}
Davide Gaiotto, Anton Kapustin, Nathan Seiberg, and Brian Willett, JHEP, {\bf 02}, 172 (2015),  {{arXiv:1412.5148}}.

\bibitem{Sharpe:2015mja}
Eric Sharpe, Fortsch. Phys., {\bf 63}, 659--682 (2015),  {{arXiv:1508.04770}}.

\bibitem{Batalin:1981jr}
I.~A. Batalin and G.~A. Vilkovisky, Phys. Lett. B, {\bf 102}, 27--31 (1981).

\bibitem{Batalin:1983ggl}
I.~A. Batalin and G.~A. Vilkovisky, Phys. Rev. D, {\bf 28}, 2567--2582, [Erratum: Phys.Rev.D 30, 508 (1984)] (1983).

\bibitem{Kawahira:2024}
Masashi Kawahira (to appear).

\bibitem{Fukushima:2022zor}
Kenji Fukushima, Takuya Shimazaki, and Yuya Tanizaki, JHEP, {\bf 04}, 040 (2022),  {{arXiv:2202.00375}}.

\bibitem{Gorantla:2021svj}
Pranay Gorantla, Ho~Tat Lam, Nathan Seiberg, and Shu-Heng Shao, J. Math. Phys., {\bf 62}(10), 102301 (2021),  {{arXiv:2103.01257}}.

\end{thebibliography}
%



\let\doi\relax

\end{document}